\documentclass[twocolumn,superscriptaddress,showpacs,nofootinbib,notitlepage,preprintnumbers,secnumarabic,amssymb, nobibnotes, aps, prl]{revtex4-2}
\usepackage[utf8]{inputenc}
\usepackage{graphicx}
\usepackage{latexsym,amsmath,amssymb,amsthm,lmodern,float,url,dcolumn}
\usepackage{natbib}
\usepackage{color}
\usepackage{microtype}
\usepackage{import}
\usepackage{bbold}
\usepackage[plain]{fancyref}
\usepackage{varioref}
\usepackage{slashed}
\usepackage{multirow}
\usepackage{tikz}
\usetikzlibrary{shapes}
\usetikzlibrary{positioning}
\usepackage[normalem]{ulem}
\usepackage{booktabs}

\usepackage[colorlinks=true,backref=false, linktocpage=true,
citecolor=blue,urlcolor=blue,linkcolor=blue,pdfpagemode=UseOutlines]{hyperref}
\hypersetup{%
  bookmarksnumbered=true,
  pdftitle = {},
  pdfsubject = {},
  pdfauthor = {},
  pdfkeywords = {}
}

\let\Re\undefined

\DeclareMathOperator{\Tr}{Tr}
\DeclareMathOperator{\Re}{Re}

\newcommand{\ds}{S({1080})}

\newcolumntype{d}[1]{D{.}{.}{#1}}

\definecolor{mycolor}{RGB}{22,139,22}

\newcommand{\redfrho}{\raisebox{0.0pt}{\tikz{\node[draw,scale=0.4,regular polygon, regular polygon sides=4,draw=red, fill=red,rotate=45](){};}}}
\newcommand{\bfcir}{\raisebox{0.8pt}{\tikz{\node[scale=0.3,circle,draw=black,fill=black](){};}}}
\newcommand{\graycir}{\raisebox{0.2pt}{\tikz{\node[scale=0.4,circle,draw=gray](){};}}}

\def\fm {\mathop{\hbox{fm}}}

\def\Re {\mathop{\hbox{Re}}}

\def\Tr {\mathop{\hbox{Tr}}}

\usepackage{amsmath} 
\usepackage{dsfont}  
\usepackage{bm}      

\def\beq{\begin{equation}}
\def\eeq{\end{equation}}
\def\beqs#1\eeqs{\beq\begin{split} #1 \end{split}\eeq}

\long\def\comment#1{}

\usepackage{mleftright,xparse}
\NewDocumentCommand\xDeclarePairedDelimiter{mmm}
{%
	\NewDocumentCommand#1{som}{%
		\IfNoValueTF{##2}
		{\IfBooleanTF{##1}{#2##3#3}{\mleft#2##3\mright#3}}
		{\mathopen{\csname##2\endcsname#2}##3\mathclose{\csname##2\endcsname#3}}%
	}%
}
\xDeclarePairedDelimiter{\av}{\langle}{\rangle}
\xDeclarePairedDelimiter{\ket}{|}{\rangle}
\xDeclarePairedDelimiter{\bra}{\langle}{|}
\NewDocumentCommand\braket{somm}{%
	\IfNoValueTF{#2}{\mleft\langle #3\,|#4\mright\rangle}{NOTIMPLEMENTED}
}
\NewDocumentCommand\opbraket{sommm}{%
	\IfNoValueTF{#2}
	{\IfBooleanTF{#1}{\langle#3|#4|#5\rangle}{\mleft\langle #3 \left| #4 \right| #5 \mright\rangle}}
	{\mathopen{\csname#2\endcsname\langle}#3\mathopen{\csname#2\endcsname|} #4 \mathclose{\csname#2\endcsname|} #5\mathclose{\csname#2\endcsname\rangle}}
}

\begin{document}
\preprint{FERMILAB-PUB-21-683-T}
\title{The spectrum of qubitized QCD: glueballs in a $\ds$ gauge theory}

\author{Andrei Alexandru}
\email{aalexan@gwu.edu}
\affiliation{Physics Department, The George Washington University, Washington, DC 20052, USA}
\affiliation{Department of Physics, University of Maryland, College Park, MD 20742, USA}
\author{Paulo F. Bedaque}
\email{bedaque@umd.edu}
\affiliation{Department of Physics, University of Maryland, College Park, MD 20742, USA}
\author{Ruair\'i Brett}
\email{rbrett@gwu.edu}
\affiliation{Physics Department, The George Washington University, Washington, DC 20052, USA}
\author{Henry Lamm}
\email{hlamm@fnal.gov}
\affiliation{Fermi National Accelerator Laboratory, Batavia,  Illinois, 60510, USA}

\date{\today}

\begin{abstract}
Quantum simulations of QCD require digitization of the infinite-dimensional gluon field. Schemes for doing this with the minimum amount of qubits are desirable. We present a practical digitization for $SU(3)$ gauge theories via its discrete subgroup $S(1080)$. Using a modified action that allows classical simulations down to $a\approx 0.08$ fm, the low-lying glueball spectrum is computed with percent-level precision at multiple lattice spacings and shown to extrapolate to the continuum limit $SU(3)$ results. This suggests that this digitization scheme is sufficient for precision quantum simulations of QCD.

\end{abstract}

\maketitle


\textit{Introduction} ---
Numerous observables remain firmly beyond the reach of numerical nonperturbative field theory~\cite{Feynman:1981tf,Jordan:2017lea,pqa_loi} due to the sign problem. Sign problems arise when the imaginary time (euclidean) action of the system is complex or when a real time representation is required.
Finite density problems (like the calculation of the equation of state of dense QCD matter or the Hubbard model away from half-filling) are famous examples of the first case; thermalization is an example of the second.

Due to the importance of these problems much effort has been spend on solving or bypassing the sign problem.
Quantum computers are a promising avenue leading to a solution to these problems. The time development of the quantum system can be directly mapped into the time evolution of the quantum computer obviating the need for imaginary time calculations.  The subtle interference patterns appearing on real time evolution are mimicked by the same pattern inside the quantum computer.

Besides the obvious technological difficulty of building quantum computers, conceptual questions must be answered before quantum field theory can be simulated. First, the state of the system needs to be mapped into a finite -- and likely small --  quantum register. Then the initial state needs to be prepared, the hamiltonian evolution coded in terms of elementary gates and, finally, observables must be measured. This paper focuses on the first step, the encoding of the states in the quantum register. The difficulty arises mainly in bosonic theories. Indeed, purely fermionic theories have an infinite dimensional Hilbert space but the usual discretization of space into a finite lattice suffices to reduce the dimensionality to a finite number mappable into a quantum register. However, the Hilbert space of a bosonic fields defined in a single lattice point is already infinite dimensional.  Thus, further discretization of {\it field space} is required to map bosons into a digital quantum register. 
 A wide array of solutions exist~\cite{Zohar:2012xf,Zohar:2013zla,Zohar:2014qma,Wiese:2014rla,Zohar:2016iic,Bender:2018rdp,Klco:2019evd,Raychowdhury:2019iki,Davoudi:2020yln,Kreshchuk:2020dla,Ciavarella:2021nmj}. Different digitizations break different symmetries of the model~\cite{Zohar:2013zla,Raychowdhury:2018osk,Wiese:2014rla,Zohar:2013zla}.  With these reductions, the universality class of the lattice model may differ from the original theory~\cite{Hasenfratz:2001iz,Caracciolo:2001jd,Hasenfratz:2000hd,PhysRevE.57.111,PhysRevE.94.022134,Alexandru:2019ozf,afmdng,Alexandru:2021xkf} making the continuum space limit problematic. Recently, studies quantified the truncation errors for quantum simulations of lattice theories from a computational complexity perspective~\cite{Davoudi:2020yln,Shaw:2020udc,Kan:2021xfc,Tong:2021rfv}.

Here, we will investigate the discrete subgroup approximation~\cite{Hackett:2018cel,Alexandru:2019nsa,Ji:2020kjk,Alam:2021uuq} using classical lattice simulations.  While performed in imaginary time, this nonperturbative study of truncation errors is known to be related to those in real time~\cite{Osterwalder:1973dx,Osterwalder:1974tc,Carena:2021ltu}, thus providing us access to much larger systems than with current quantum devices.  Discrete subgroups were studied in the early days of lattice field theory when memory limitations restricted the feasible lattice volumes due to the cost of storing $SU(3)$ elements.  

Replacing a continuous symmetry with a discrete (and smaller)
may easily destroy the proper (spacetime) continuum limit.
 In asymptotically free theories like QCD, the continuum limit is obtained as the inverse coupling $\beta$ diverges while the lattice spacing~$a$ goes to zero. However, for~$a$ smaller than a certain threshold~$a_f$ the discrete theory differs drastically from the one with a continuous group~(although there are counterexamples~\cite{Beard:2004jr}).  In the language of euclidean path integrals, the field configurations are ``frozen" on the configurations with the minimal action, with the other configurations, due to the gap in action, being  exponentially suppressed.  This freezing is not necessarily fatal provided we reach the scaling regime: realistic lattice calculations are performed on classical computers with a finite $a$ and extrapolated to $a\rightarrow0$. In these calculations, $a$ should be smaller than typical hadronic scales with modern values of $\mathcal{O}(0.1$~fm).
 
 Subgroups of $U(1)$~\cite{Creutz:1979zg,Creutz:1982dn,Fukugita:1982kk} and $SU(N)$~\cite{Petcher:1980cq,Jacobs:1980mk,Bhanot:1981xp,Grosse:1981bv,Bhanot:1981pj,Lisboa:1982jj,Flyvbjerg:1984dj,Flyvbjerg:1984ji} gauge fields -- including with fermions~\cite{Weingarten:1980hx,Weingarten:1981jy} -- were tested with differing degrees of success. In particular, all five crystal-like subgroups of $SU(3)$ freeze before the scaling regime with the Wilson action~\cite{Bhanot:1981xp,Bhanot:1981pj,Alexandru:2019nsa}. Subsequent work increased the phase transition by including the midpoints between elements of $\ds$ ~\cite{Lisboa:1982jj}. However, this procedure breaks gauge symmetry. An alternative proposal studied for~$U(1)$ chooses an optimized subset of group elements~\cite{Haase:2020kaj,Bauer:2021gek}.


Introducing new terms to the discrete action can decrease $a_f$. 
The new action has formally the same continuum limit, but it is different at finite lattice spacing.
The net effect is that the gap $\delta S$ between the lowest action configurations is lowered and the freezing $a_f$ is reduced. Classical Monte Carlo calculations were performed in~\cite{Alexandru:2019nsa}, finding that $a\approx 0.08$~fm is possible with the addition of a single term to the action. We must check empirically that this action generates the same physics as $SU(3)$ by reproducing continuum IR observables. The previous work~\cite{Alexandru:2019nsa} demonstrated that the deconfining temperature of pure-glue $SU(3)$ was reproduced in the continuum from $\ds$ to sub-percent accuracy. 
In order to argue for the usefulness of this approach to quantum simulations we need verify that the $S(1080)$ is capable of reproducing the hadronic spectrum.
Here, we show that percent-level accuracy can be achieved for the low-lying glueball states, the massive excitations of pure-glue SU(3).

\textit{Theory} ---
 In this work, we use the action proposed in~\cite{Alexandru:2019nsa}:
\beqs
 S[U]&=-\sum_p \left(\frac{\beta_0}{3}\Re\Tr (U_p) +\beta_1\Re\Tr(U_p^2)\right),
\label{eq:action-mod}
\eeqs
where $U_p$ is a plaquette and $\beta_i$ are coupling constants. Modified actions like Eq.~(\ref{eq:action-mod}) were observed for $U_p\in SU(3)$ to have milder lattice spacing errors~\cite{Blum:1994xb,Heller:1995bz,Heller:1995hh,Hasenbusch:2004wu,Hasenbusch:2004yq} with the same continuum limit as the Wilson action ($S_{W} = -\tfrac{\beta_{W}}{3} \sum_p \Re\Tr U_p$). 

The inability of the discrete field fluctuations to be made arbitrarily small lead to a phase boundary $\{\beta_i\}$ beyond which $\ds$ gauge links become frozen to $\mathbb{1}$ while the $SU(3)$ gauge links remain dynamical. In this phase, no clear connection between $\ds$ and $SU(3)$ exists. That is what limits the lattice spacing to $a>a_f$.  By setting $\beta_1<0$ the gap between the values of the frozen and dynamical discrete links is reduced and thus $a_f$ decreased.

In~\cite{Alexandru:2019nsa} the trajectory $\beta_1=-0.1267\beta_0+0.253$ was found to allow for lattice spacing down to $a\gtrsim0.08$ fm when computing the Wilson flow parameter $t_0$~\cite{Luscher:2010iy}.  This trajectory was chosen to avoid the freezing transition in a small $2^4$ lattice.  This value of $a\gtrsim 0.08$ fm is clearly in the scaling regime where the correlation lengths are much smaller than $a$ and lattice artifacts are small.  

\begin{figure}[t]
 \begin{center}
  \includegraphics[width=0.9\columnwidth]{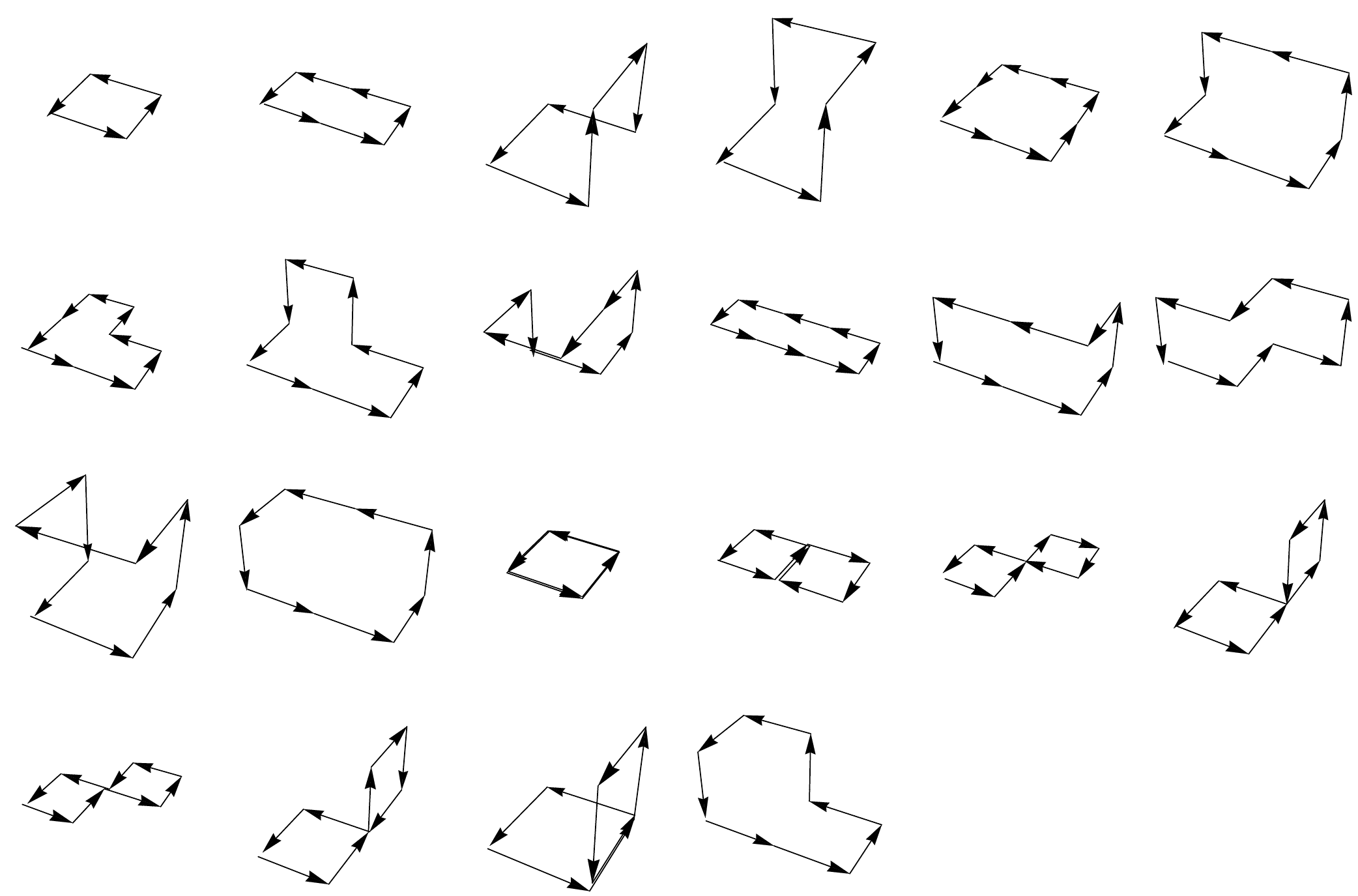}
 \end{center}
\caption{\label{fig:loops}Seed loops for the operator
constructions, using 4, 6, and 8 links.}
\end{figure}

\textit{Methodology} ---
To extract glueball masses we need to measure two-point functions
of operators with the appropriate quantum numbers. 
Gluon correlators are noisy and the extraction of gluon masses hinge on a good choice of interpolating fields and a
 a variational calculation employing a large set
of operators. Fortunately, a sophisticated technique has been developed that we can use for glueball spectroscopy~\cite{Berg:1982kp,Berg:1983qd,Morningstar:1999rf,Wenger:2000,Chen:2005mg,Athenodorou:2020ani}.
While the traditional approach for $SU(3)$
glueball spectroscopy involves anisotropic lattices~\cite{Morningstar:1999rf}, we used isotropic lattices, to avoid the complication of tuning the anisotropy.
These operators
need to be gauge invariant and are constructed from traces of 
{\em loops}, sets of links that track paths that return to the
starting point. The basic seed paths we used for this study
are one 4-link long, three 6-link long, and 18 8-links long (Fig.~\ref{fig:loops}). Generically an $n$-link loop operator is given by
\beq\label{eq:loops}
[x; \mu_1,\ldots,\mu_n] \equiv \Tr \prod_{i=1}^n U_{\mu_i}\Big(x+\sum_{j<i} \mu_j\Big) \,.
\eeq
Above, $x$ is the starting point for the loop and $\mu_i$ are 
{\em spatial} displacements for the loop. Since this is a loop, 
the displacements satisfy $\sum_k \mu_k=0$. The {\em links}
$U_\mu(x)$ represent a Wilson line connecting $x$ and $x+\mu$.
As the lattice spacing is reduced the loops are enlarged by {\em blocking}, 
 that is, repeating the steps in the loop, so that the size of the loops in physical units stays constant.
For example $[\mu_1,\mu_2,\ldots,\mu_n]$
can be replaced with $[\mu_1,\mu_1,\mu_2,\mu_2,\ldots,\mu_n,\mu_n]$.

We will consider only zero-momentum operators
resulting from summing \eqref{eq:loops} over $x$.
These operators have symmetries that we use to make
their calculation more efficient: 
they are invariant under circular permutations of the steps, and the opposite orientation loop
is related to the original one via a complex-conjugation.


The loop operators must be projected onto the appropriate irreducible
representations ({\em irreps}) of the finite-volume symmetry 
group~$O_h$ (we consider only cubic boxes),
a subgroup of the infinite-volume symmetry group, $O(3)$.
These projectors are given by:
\beq
P^\Gamma_{\lambda\lambda'} \ell \equiv
\frac{d_\Gamma}{|G|} \sum_{g\in G} [D^\Gamma_{\lambda\lambda'}(g)]^* R(g)\ell \,, 
\eeq
where $\ell=[\mu_1,\ldots,\mu_n]$ and $R\ell=[ R\mu_1,\ldots,R\mu_n]$ is the loop transformed by
$R$. Above, $d_\Gamma$ is the dimension
of the irrep $\Gamma$, $|G|$ is the number of elements in the
group, $D^\Gamma(g)$ is the matrix associated with element
$g$ in irrep $\Gamma$ and $R(g)$ is the three-dimensional
representation of $O_h$ (these are the same matrices as
$D^{T_1^-}(g)$.) The group $O_h$ has 24 proper rotations,
and 24 rotations combined with an inversion. There
are 10 irreps: $A_1^{\pm}$, $A_2^\pm$, $E^\pm$, $T_1^\pm$,
and $T_2^\pm$ with dimensions 1, 1, 2, 3, and 3.
The multiplets of $O(3)$  can be decomposed into smaller invariant 
multiplets  of $O_h$. For instance, the scalar ($J=0$) irrep corresponds to the $A_1$ irrep of $O_h$ while the $J=2$ breaks down into $E \oplus T_2$. 
The parity  are the same for both 
infinite volume and finite volume irreps. 


The other quantum number relevant for operator construction is the charge parity. 
Charge conjugation for the glue fields is given by $U_\mu(x)\to U_\mu(x)^*$,
so the loop operators transform similarly $\ell\to \ell^*$. The even-charge
operators, whiche we consider here, correspond to the real part of $\ell$ and the odd ones to the imaginary parts.
Finally, in order to increase the overlap of the loop-operators with the glueball states
we {\em stout smear} the links~\cite{Morningstar:2003gk}.


In our calculation, we used the following operator basis. All loops
were computed using smeared operator on blocked links. We used 16 different 
combinations. For each gauge-configuration we repeatedly smeared the links
generating four different smearing levels with $n_\text{smear}=2,4,6,8$. 
For each smearing set of links we evaluated the loops
using blocked links with $n_\text{block}=2,4,6,8$.

Since the gluon correlators becomes noisy at fairly small time separations, a delicate fitting procedure is required to extract the glueball masses.
Finite-volume glueball energies are best extracted by computing {\emph matrices} of temporal correlators,
\beq
C_{ij}(\tau) = \sum_{\tau_0} \langle 0 |{\mathcal{O}_i(\tau+\tau_0){\mathcal{O}}_j(\tau_0)}^\dagger|0\rangle,
\eeq
for large sets of glueball operators $\mathcal{O}(\tau)=O(\tau)-\langle 0|O(\tau)|0\rangle$. In practice, the vacuum subtraction only needs to be performed for operators with vacuum quantum numbers, i.e.~the $A_{1}^{++}$ sector. We construct 
the matrix
\beq
\widetilde{C}(\tau) = U^\dagger C(\tau_0)^{-1/2} C(\tau) C(\tau_0)^{-1/2} U,
\eeq
where the columns of $U$ are the eigenvectors of $G(\tau_d)=C(\tau_0)^{-1/2} C(\tau_d) C(\tau_0)^{-1/2}$. 
The parameters $\tau_0$ and $\tau_d$, named the pivot and diagonalization times, are chosen such that $\widetilde{C}(\tau)$ remains (approximately) diagonal for $\tau>\tau_d$, and the extracted energies are insensitive to parameter variations.
The spectrum is then extracted by fitting the diagonal elements $\widetilde{C}_{kk}(\tau)$ to the ansatz 
$A_k[e^{-E_k \tau} + e^{-E_k(T-\tau)}]$,
with $T$ the extent of the Euclidean lattice time.
The ground state is associated with the largest eigenvalue of $G(\tau_d)$, the first excited state with the second largest eigenvalue, and so on.

From the 22 seed loops in Fig.~\ref{fig:loops}, 626 linearly independent operators are produced across the 20 $\Gamma^C$ symmetry sectors. 
Performing the smearing and blocking process leads to 10,016 independent operators for the 16 sets of links.
While the full set of operators for a given irrep can be used, in practice it is often necessary to carefully prune the operator basis.
This is done by removing operators with poor overlap onto the states of interest, and those whose correlators have a low signal-to-noise ratio.
Unlike modern QCD calculations~\cite{Brett:2019tzr,Culver:2019vvu} here we are only interested in the ground state in a given irrep.
Therefore, while pruning is helpful in simplifying the analysis to smaller matrices, we find our results insensitive to it.

\begin{figure}[h!]
 \begin{center}
  \includegraphics[width=\linewidth,trim=0 1.5cm 0 0.5cm,clip]{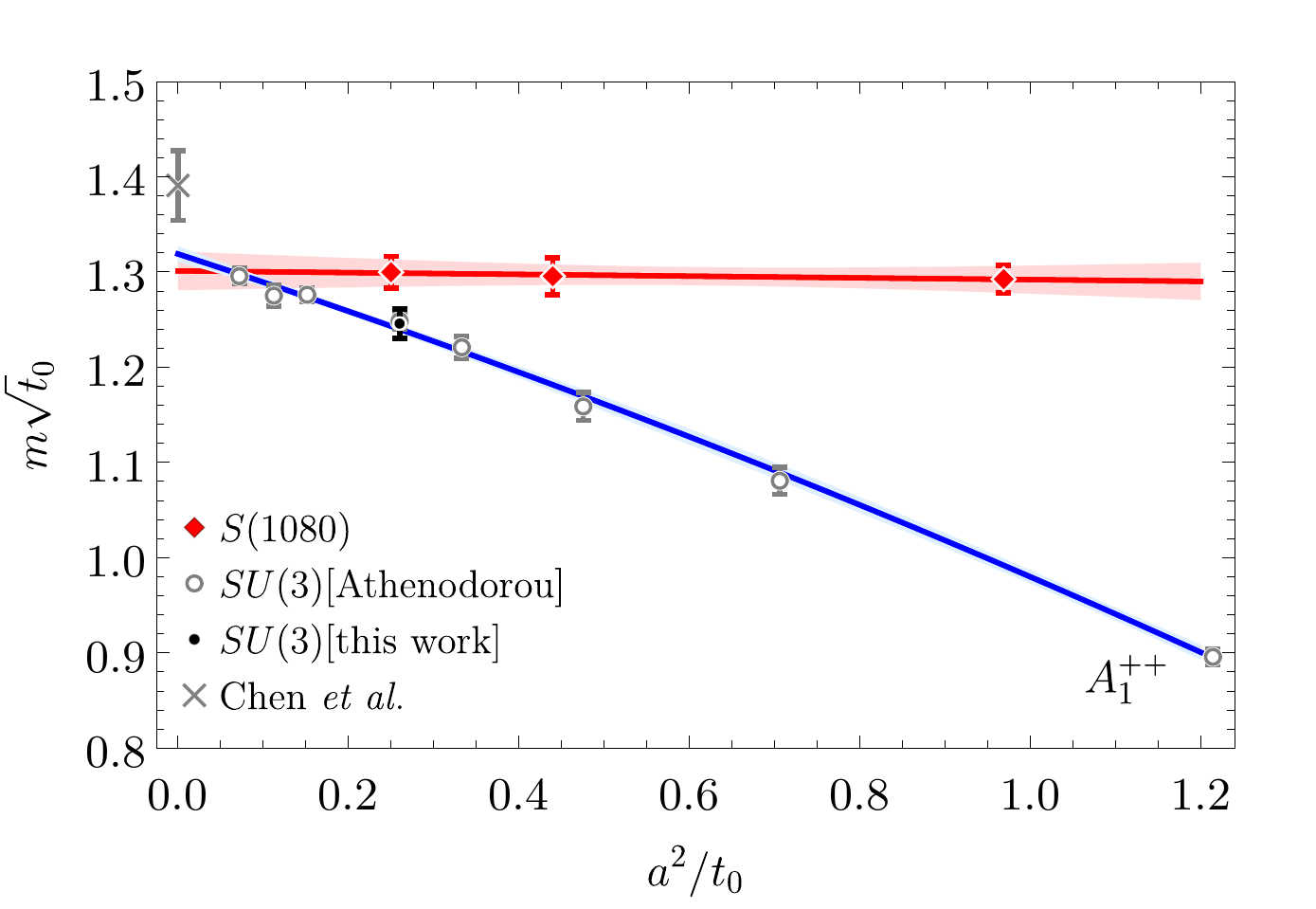}
  \includegraphics[width=\linewidth,trim=0 1.5cm 0 0.5cm,clip]{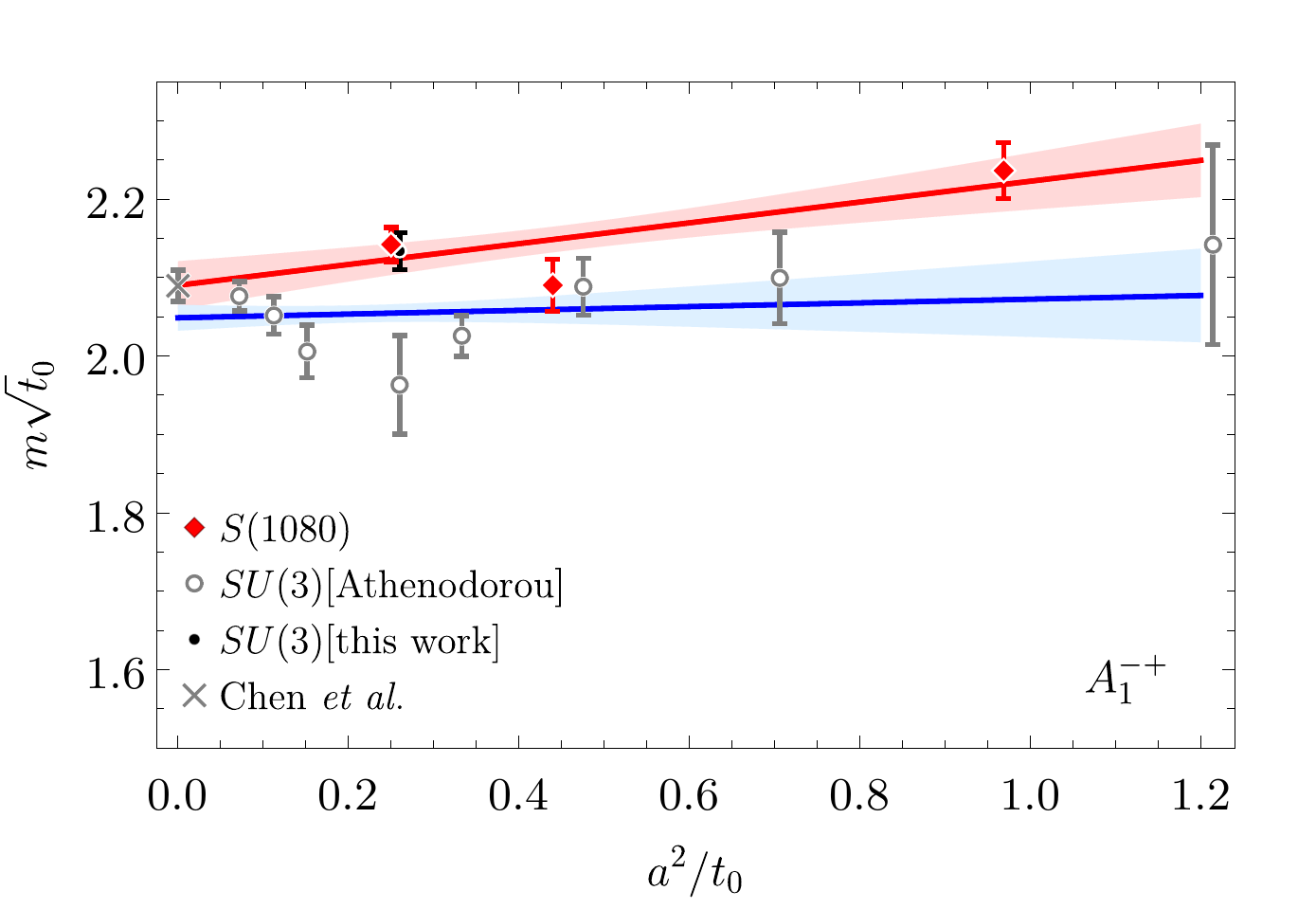}
  \includegraphics[width=\linewidth,trim=0 0cm 0 0.5cm,clip]{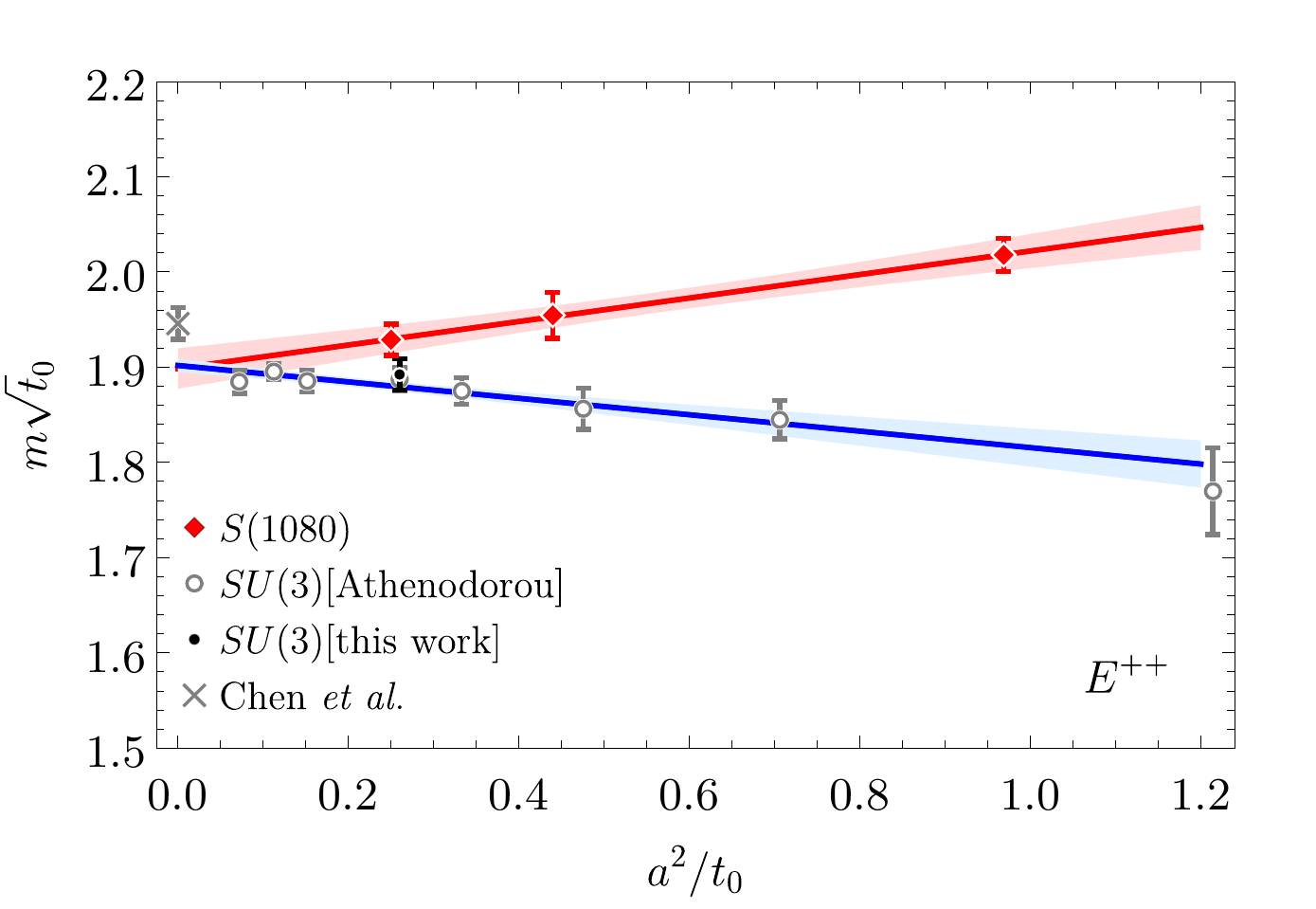}
 \end{center}
\caption{\label{fig:masses} Masses for $A_1^{++}$ ($J^{PC}=0^{++}$), $A_1^{-+}$ ($J^{PC}=0^{-+}$), and $E^{++}$ ($J^{PC}=2^{++}$) glueballs vs $a^2/t_0$.  Our $\ds$ (\protect\redfrho) and $SU(3)$ (\protect\bfcir) results compared to $SU(3)$ results from~\cite{Athenodorou:2020ani}~(\protect\graycir). Another continuum $SU(3)$ result~\cite{Chen:2005mg}~($\times$) is presented to estimate systematic errors.}
\end{figure}

\textit{Results} ---
Our results are obtained from three $\ds$ ensembles using the same couplings as~\cite{Alexandru:2019nsa}. The parameters were chosen to scan a set $a\in[0.08, 0.16]\fm$. The lattice spacing was determined from $t_0$, the Wilson flow time~\cite{Luscher:2010iy}. A detailed list of parameters is in Table~\ref{tab:s1080params}. These ensembles were generated using a multi-hit Metropolis update algorithm, which we found to be as efficient as a heat-bath in terms of autocorrelation length but significantly cheaper to implement. For each ensemble we generated around 650,000 independent configurations with sufficient decorrelation steps. Computing time was dominated by measuring the loop-operators, so we used a conservative number of update steps between measurements.

\begin{table*}
\caption{Input parameters. The top three lines are for $\ds$ and the forth is the $SU(3)$ calibration run. The parameters are: $\rho$ the stout smearing parameter, $n_{\rm decorr}$ the number of updates between measurements, $n_\rho$ and $n_b$ the number of smearing and blocking levels respectively. For $\ds$, the couplings $\beta_0$ and $\beta_1$ are normalized as in Eq.~(\ref{eq:action-mod}), and follow the trajectory in the text, whereas for $SU(3)$ simulations $\beta_0$ is the Wilson coupling. Observables are quoted in lattice units, with statistical errors only. For $SU(3)$ the value of $\sqrt{t_0}/a$ is from~\cite{Francis:2015lha}.}
\label{tab:s1080params}
\centering
\begin{tabular*}{0.99\textwidth}{@{\kern1ex\extracolsep{\stretch{1}}} r  c c c  c  c  c  c  c  c  c l  l  l @{\kern1ex}}
    \toprule
    \multicolumn{1}{c}{$\beta_0$} & $\beta_1$ & \phantom{a} &
     $n_x^3 \times n_t$ & $n_{\rm therm}$ & $n_{\rm decorr}$ & $\rho$ &
    $n_{\rm meas}$ & $n_{\rm bins}$ & 
    \phantom{a} &
    $\sqrt{t_0}/a$ &
    \multicolumn{1}{c}{$a m_{A_1^{++}}$} & \multicolumn{1}{c}{$am_{A_1^{-+}}$} & \multicolumn{1}{c}{$am_{E^{++}}$} \\
    \midrule
    $9.154$ & $-0.9065$ && $16^3\times16$ & $200$ & 40 & $0.2$ & 
    652500 & 1305 && $1.016(3)$ & 1.272(14) & 2.201(35) & 1.986(17) \\
    $12.795$ & $-1.3677$ && $16^3\times16$ & $200$ & 40 & $0.2$ & 
    650000 & 1300 && $1.508(3)$ & 0.859(13) & 1.386(22) & 1.296(16) \\
    $19.61$\phantom{0} & $-2.2309$ && $16^3\times16$ & $200$ & 40 & $0.2$ &
    647500 & 1295 && $2.000(4)$ &  0.6498(84) & 1.071(11) & 0.9644(82) \\
    \midrule
    $6.0625$ & --- && $16^3\times16$ & $200$ & 5 & $0.2$ &
    567500 & 1135 && 1.962(1) & 0.6352(79) & 1.088(12) & 0.9649(84) \\
    \bottomrule
\end{tabular*}
\end{table*}

To verify that our operator basis overlaps well with the glueballs and that our fitting method is sound, we carried out a calibration run using an $SU(3)$ ensemble. We generated a set of $SU(3)$ Wilson-gauge configurations for $\beta=6.0625$ and compare our results with the state-of-the-art calculations of~\cite{Athenodorou:2020ani} for one value of $\beta$ that is near the lattice spacing on one of our $\ds$ ensembles. We generated similar statistics to the $\ds$ ensembles and used the same set of operators. The full set of parameters for this run is included in the last row of Table~\ref{tab:s1080params}. 

For analysis we binned the measurements in groups of 500, both to remove the possible autocorrelation effects and to make the data analysis more manageable.
We fit a single exponential function, $A[e^{-m \tau} + e^{-m(T-\tau)}]$, to the ground state correlator in the time range $\tau\in[\tau_i,\tau_f]$. The relevant parameters for these calculations are in Table~\ref{tab:fitparams}. The masses and their errors were extracted using a correlated fit to take into account the covariance of the correlator at different times. The covariance matrix for the correlator was estimated using the jackknife method. The results are plotted in Fig.~\ref{fig:masses} and included in Table~\ref{tab:s1080params}.

\begin{table}[ht]
    \centering
    \caption{\label{tab:fitparams} Analysis parameters for the $\ds$ calculation (top rows) and the $SU(3)$ calibration run (bottom). Included are the  pivot time $\tau_0$, the diagonalization time $\tau_d$, the fit range $[\tau_i,\tau_f]$, and the $\chi^2$ per degree of freedom. }
    \begin{tabular*}{0.99\columnwidth}{@{\kern1ex\extracolsep{\stretch{1}}} c c c c c c c @{\kern1ex}}
    \toprule
        $\beta_0$ &  irrep & $\tau_0$ & $\tau_d$ & $\tau_i$ & $\tau_f$ & $\chi^2$/dof \\
    \midrule
        \phantom{0}
        9.154 & $A_1^{++}$ & 0 & 1 & 2 & 7 & 0.94 \\
              & $A_1^{-+}$ & 1 & 2 & 1 & 15 & 0.82 \\
              & $E^{++}$   & 0 & 1 & 1 & 15 & 1.37 \\
        12.795& $A_1^{++}$ & 0 & 1 & 3 & 8 & 1.24 \\
              & $A_1^{-+}$ & 0 & 1 & 2 & 7 & 0.96 \\
              & $E^{++}$   & 0 & 1 & 2 & 8 & 1.91 \\
        19.61\phantom{0}
              & $A_1^{++}$ & 0 & 1 & 3 & 8 & 1.47 \\
              & $A_1^{-+}$ & 0 & 1 & 2 & 5 & 1.11 \\
              & $E^{++}$   & 0 & 1 & 2 & 8 & 0.83 \\
        \midrule
        6.0625& $A_1^{++}$ & 2 & 3 & 3 & 10 & 1.01 \\
              & $A_1^{-+}$ & 0 & 1 & 2 & 7 & 0.90 \\
              & $E^{++}$   & 1 & 2 & 2 & 8 & 1.07 \\
    \bottomrule
    \end{tabular*}
\end{table}

We extracted the glueball masses corresponding to the ground states in the $A_1^{++}$, $E^{++}$, and $A_1^{-+}$ irreps, corresponding to the lowest lying glueballs. 
The $SU(3$) calibration run results are included in Fig.~\ref{fig:masses} and are consistent with those from Ref.~\cite{Athenodorou:2020ani}. The $A_1^{-+}$ point differs from the corresponding mass from Ref.~\cite{Athenodorou:2020ani}, yet we note that at this point the results from Ref.~\cite{Athenodorou:2020ani} have large error-bars and are at tension with their own continuum extrapolation.
Our $\ds$ results are extrapolated to $a=0$ assuming the expected quadratic form $m(a)\sqrt{t_0}=m(0)\sqrt{t_0}+c a^2/t_0$. These extrapolations are indicated in Fig.~\ref{fig:masses} with a red line and compared with the $SU(3)$ extrapolations from Refs.~\cite{Athenodorou:2020ani,Chen:2005mg}. For our calculations we use $\sqrt{t_0}/a$ values measured directly on these ensembles~\cite{Alexandru:2019nsa} (see Table~\ref{tab:s1080params}). For the $SU(3)$ results we used the parameterization of $\sqrt{t_0}/a$ as a function of $\beta$ for the pure glue Wilson action included in a recent study~\cite{Francis:2015lha} and we perform the same continuum extrapolation as in Ref.~\cite{Athenodorou:2020ani}.
To gauge the systematics of the $SU(3)$ calculation, we included the results from an independent calculation~\cite{Chen:2005mg}. The extrapolation results are included in Table~\ref{tab:extrapolation}. As we can see, the results agree within their statistical errors at the percent level. Compared to previous results for the deconfining temperature $T_0\sqrt{t_0}=0.2489(11)$, our results probe nearly an order of magnitude higher in energy $m\sqrt{t_0}\sim2$ finding agreement with $SU(3)$ in the continuum. This  supports the claim that this modified action reproduces $SU(3)$ physics below 2.5 GeV$^{-1}$.

\begin{table}[ht]
    \centering
    \caption{\label{tab:extrapolation} $a\rightarrow0$ extrapolations of $m\sqrt{t_0}$, for $\ds$ and $SU(3)$ simulations. The results of~\cite{Athenodorou:2020ani} uses an isotropic lattice whereas \cite{Chen:2005mg} use an anisotropic lattice.}
    \begin{tabular*}{0.99\columnwidth}{@{\kern1ex\extracolsep{\stretch{1}}} c c c c  @{\kern1ex}}
    \toprule
         irrep & $\ds$ & $SU(3)$ \cite{Athenodorou:2020ani}  & $SU(3)$ \cite{Chen:2005mg}\\
         \midrule
         $A_1^{++}$ & 1.301(20) & 1.319(8)\phantom{0} & 1.391(37)\\
         $A_1^{-+}$ & 2.090(31) & 2.049(17) & 2.089(20)\\
         $E^{++}$ & 1.899(21) & 1.902(7)\phantom{0} & 1.946(17) \\
         \bottomrule
    \end{tabular*}
\end{table}

\textit{Conclusions} ---
These results provides strong evidence that $\ds$ can replace $SU(3)$ in quantum simulations of observables which cannot be computed using classical lattice techniques in imaginary time e.g.~\cite{Cohen:2021imf}. This includes some key QCD phenomena that have remained mysterious up to now like the mechanism of thermalization in heavy ion collisions that is believed to be driven mostly by gluons.
The $\mathcal{O}(10^2)$ qubit savings from using an 11-qubit $\ds$ register instead of $SU(3)$ compare favorably to other digitizations~\cite{Banerjee:2012pg,Banerjee:2012xg,Marcos:2014lda,Rico:2018pas,Ciavarella:2021nmj}. Even with the small lattice sizes dominating the error due to the limited number of qubits, the model studied here represents a sufficient approximation of $SU(3)$ for $a\gtrsim 0.08$ fm. As quantum computers get larger, smaller lattice errors may become desirable.  In such case, systematic improvements are possible. Including additional terms in the action proportional to other characters~\cite{Flyvbjerg:1984dj,Flyvbjerg:1984ji,Ji:2020kjk} can allow for smaller $a$, while improved actions, in the spirit of the Symanzik program,  can reduce the systemic error for fixed $a$~\cite{Symanzik:1983dc,Luscher:1984xn,Luscher:1985zq}.  The relative cost of these two improvements are left for future work. Finally, including dynamical fermions into discrete subgroups simulations should be performed to understand how this full theory compared to QCD. While there is no conceptual issue in including fermions, the numerical value of the minimum lattice size achievable will depend on the number of dynamical quarks.


\begin{acknowledgments}

A.A. and R.B. are grateful to Andreas Athenodorou for his 
guidance on constructing a good operator basis for the 
variational analysis.
P.B was supported in part by the US DoE under contract No. DE-FG02-93ER-40762 and by U.S. DOE Grant No. DE-SC0021143.
 A.A. is supported in part by the U.S. Department of Energy grant DE-FG02-95ER40907. 
H.L is supported by the Department of Energy through the Fermilab QuantiSED program in the area of ``Intersections of QIS and Theoretical Particle Physics". Fermilab is operated by Fermi Research Alliance, LLC under contract number DE-AC02-07CH11359 with the United States Department of Energy. 
\end{acknowledgments}

\bibliography{refs}

\end{document}